# Spectroscopy of Single CdSe Magic-Sized Nanocrystals

*Gabriel Nagamine, Julian Santen, Juri G. Crimmann, Aniket S. Mule, Andrew B. Pun, and David J. Norris*[*]

Optical Materials Engineering Laboratory, Department of Mechanical and Process Engineering,

ETH Zurich, 8092 Zurich, Switzerland

ABSTRACT: Chemical syntheses that provide nanocrystals (NCs) with narrow distributions in size and shape are critical for NC research. This has led to the investigation of magic-sized NCs (MSNCs), a class of semiconductor crystallites that grow in discrete steps, potentially offering a single size and shape (*i.e.*, monodispersity). However, the photoluminescence (PL) spectra of CdSe MSNCs measured at room temperature have been reported to be broader than those of state-of-the-art quantum dots. This difference could be due to the smaller size of MSNCs, which broadens their line widths, or due to their residual size dispersity. To better understand the optical performance of MSNCs, here we perform single-particle spectroscopy. Our results show that, while CdSe MSNCs do exhibit particle-to-particle variations that lead to modest broadening of their ensemble emission spectra, the largest contribution comes from the single-particle line width. By examining MSNCs with different sizes and shells, we conclude that this single-particle broadening is consistent with exciton coupling to acoustic phonons from the NC surface. Because of their small size, this coupling and the role of residual size dispersity have a larger impact on the ensemble emission line widths. Notably, when small (<2.7 nm diameter) MSNCs and quantum dots are compared, the ensemble PL line widths of MSNCs are actually sharper. Due to their small size, MSNCs also exhibit strong anti-bunching [$g^{(2)}(0) \sim 0.05$] at room temperature. Thus, MSNCs represent a bright, spectrally pure class of quantum emitter, useful for applications in optoelectronic and quantum-information technologies where strong three-dimensional confinement is required.

INTRODUCTION

The preparation of colloidal semiconductor nanocrystals (NCs) that are all exactly the same size and shape (*i.e.*, true monodispersity) through wet-chemical synthesis has been a long-standing goal.[1-5] Our ability to control NC size and shape has major implications for the applications of these materials. Narrow size distributions allow sharper emission line widths, which enhances color purity for display technologies.[6,7] Moreover, color-pure NCs are potentially useful for quantum information, in which many identical single-photon sources are required.[8,9] However, the synthesis of NCs with true monodispersity is impeded by the NC growth mechanism. Conventional wet-chemical protocols involve the continuous attachment of atoms onto the NC surface. Due to the randomness of this process, one would expect to obtain NC samples with a range of sizes and shapes. Indeed, even state-of-the-art syntheses of quasi-spherical NCs (known as quantum dots) and cube-shaped NCs present size distributions of ~5%.[10-12]

To avoid such distributions, researchers have explored whether an alternative growth strategy can get closer to monodispersity. Instead of continuous addition of material, several synthetic procedures have been discovered that lead to discrete jumps in size. This "discrete growth" has been observed in two classes of semiconductor nanomaterials: (i) nanoplatelets (NPLs)[13-15] and (ii) magic-sized clusters (MSCs).[16-18] NPLs are thin rectangular crystallites[19,20] that are 10's to 100's of nanometers in plane but have a thickness of only a few atomic layers. This thickness dictates their optical properties (*e.g.*, emission and absorption spectra). Amazingly, synthetic procedures have been developed that can yield samples in which all NPLs share the same thickness (*e.g.*, 4 atomic layers).[19] Thus, NPLs are truly monodisperse in one dimension and present sharper emission line widths than conventional quantum dots (QDs).[6] Moreover, during growth, NPL thickness slowly increases in discrete jumps (*e.g.*, 2 to 3 to 4 atomic layers), leading to a spectral series of narrow emitters.[19]

We recently proposed[21,22] that NPLs are strongly connected to magic-sized clusters[23,24]—the second class of discretely growing NCs. MSCs are extremely small crystallites containing a countable number of atoms that appear very early in some NC syntheses. They arise because they have a "magic" atomic arrangement that is more stable than crystallites that are slightly smaller or larger. Moreover, during growth, they exhibit discrete jumps between a series of such MSCs.[21-23,25] For the smallest of these, single-crystal X-ray diffraction (XRD) has even been exploited to extract their atomic structure,[26,27] suggesting (near) monodispersity. Perhaps more surprising, conditions have also been discovered that extend the discrete growth of MSCs to much larger sizes (>3 nm), beyond what is typically thought of as the "cluster" regime. For example, we previously reported a synthetic protocol for CdSe NCs that produces 10 discrete jumps.[21] We refer to the entire series of 11 sizes (from small MSCs to large discretely growing NCs) as magic-sized nanocrystals (MSNCs). While MSNCs have often suffered previously from low band-edge photoluminescence (PL) quantum yields (QYs)[28] and/or strong "deep trap" emission,[23,29] MSNCs are now available with band-edge PL QYs as high as 50% for core-only MSNCs[21] and up to 80% if a shell is added.[30] Thus, MSNCs represent a colloidal nanomaterial that is strongly emitting while potentially providing monodispersity in all three dimensions due to their discrete growth.

If MSNCs are truly monodisperse, their emission lines would presumably be narrow. Table 1 compares reported room-temperature emission spectra of ensembles of CdSe MSNCs[21] with those of other types of colloidal CdSe-based NCs.[19,31-33] The listed materials exhibit different levels of quantum confinement, which can certainly affect the emission width.[34] However, the observed lines of MSNCs appear broader than NPLs as well as state-of-the-art QDs, either with or without a shell. Since the main goal of producing discretely grown NCs is to obtain size monodispersity, ideally with spectrally pure emission, these observations then lead to several important questions: (i) How large are particle-to-particle variations in the emission spectra of individual MSNCs? (ii) What is the emission line width of a single MSNC? (iii) What are the main processes responsible for the spectral broadening in the emission spectra of MSNCs? (iv) If quantum confinement is considered, how do MSNCs perform compared to other state-of-the-art NCs?

To address these questions, here we perform single-NC spectroscopy[35] on CdSe MSNCs. Looking at individual particles allows us to access information on the particle-to-particle variations and their intrinsic single-particle line widths.[35,36] After isolating one discrete size of MSNC, we study differences between the ensemble and single-particle behavior. We find that the major contributor to the observed PL broadening is the single-particle emission line width. To gain insights into the physical mechanisms responsible for this broadening, we then investigate the influence of size and surface passivation on the emission spectra and lifetimes. We see evidence that the emission broadening is driven by enhanced exciton coupling to acoustic phonons, which for MSNCs is higher than commonly studied CdSe NCs due to stronger confinement in three dimensions (3D). This indicates that MSNCs of larger size and improved surface passivation may enable narrower line widths. More generally, when comparing emission lines between different NCs (as in Table 1), one must consider the confinement. Stronger 3D confinement leads to broader single-particle emission due to enhanced phonon coupling.

We find that the emission line widths of small MSNCs are actually narrower than the best-reported QDs of similar size.

**RESULTS AND DISCUSSION**

**Particle-to-Particle Variations. Figure**a illustrates two different sources of line-width broadening in NCs. The ensemble PL results from the collective behavior of many individual NCs. Therefore, it is necessary to separate contributions from individual particles from the variations among particles to understand the physical mechanisms responsible for PL line widths.

To achieve this separation, we performed single-particle spectroscopy[35] on MSNCs (see Methods) at room temperature. **Figure**b–d summarizes results for a specific size of CdSe MSNC. Specifically, the sample has its lowest-energy absorption feature at 494 nm (CdSe$_{494}$, Figure S1 in the Supporting Information). Figure 1b compares the ensemble PL spectrum with that of a representative individual MSNC. The ensemble emission has a line width [full-width at half-maximum (FWHM)] of 122 meV and is centered at 2.43 eV. The single-particle emission has an FWHM of 97 meV and is also centered at 2.43 eV. (All values were obtained by fitting the PL spectra to Gaussian functions.) The sharper line width observed in the single-particle PL indicates that particle-to-particle emission variations are present in the studied MSNCs. To quantify these variations, we measured single-particle emission spectra of 194 CdSe$_{494}$ MSNCs. To verify the representability of these particles, we averaged the spectra of all single MSNCs and compared them with the ensemble PL (Figure S2). From this comparison, we confirmed that the resulting line width obtained by summing the individual particles was consistent with the ensemble PL line width (see Section S1 in the Supporting Information for further details).

Figure 1c,d summarizes the particle-to-particle variations observed in the single-MSNC experiments. Figure 1c shows the distribution of center emission energies for the analyzed particles, with a standard deviation of 32 meV, which is 26% of the ensemble PL FWHM of 122 meV. Variations in the single-MSNC line width (FWHM) are also observed, with a standard deviation of 11 meV (**Figure**d). Moreover, the average single-particle FWHM was 98 meV, 22% sharper than the ensemble FWHM. These results indicate that particle-to-particle emission variations are still present in our MSNCs and contribute to the ensemble PL line width.

We now consider these results in light of the MSNC growth model recently proposed by our group.[21] It states that MSNCs are consistent with layer-by-layer growth on tetrahedron-shaped NCs. The addition of a new layer on any one of the four identical facets of the tetrahedron creates the next larger size in a series of tetrahedra. While this model explains the discrete growth of a series of MSNC sizes, it cannot rule out the possibility of morphological variations between tetrahedra of the same nominal size. For instance, atoms from the edges or vertices might be missing, leading to truncated tetrahedra. Indeed, morphological differences are observed in transmission electron microscopy (TEM) images of CdSe MSNCs.[21] While many particles exhibit triangular two-dimensional (2D) projections in such images (as expected for tetrahedra), others show irregular projections indicative of truncated tetrahedra. The variations observed in PL in Figure 1c,d are consistent with these observations. (It remains unclear when these truncations occur, during the growth or the subsequent isolation, sample preparation, or optical experiments.)

Although we observed particle-to-particle emission variations on our MSNCs, the main contributor to the ensemble PL broadening is the single-particle line width, which, on average, was 98 meV. This value represents a fundamental limit for the achievable line width of our MSNCs.

**Origin of the Single-Particle Line Width.** To investigate the origin of the observed single-MSNC line widths, we start by looking at the influence of size. To achieve a better comparison with QDs, we grew larger MSNCs. Synthesizing large MSNCs can be challenging since the time required to obtain the next discrete size increases exponentially as the particles grow.[21,22,25] But recently, a strategy was reported that accelerates the growth kinetics, leading to a CdSe MSNC with its lowest-energy absorption peak at 565 nm (CdSe$_{565}$, see Figure S1).[21]

Figure a compares the ensemble line width of this CdSe$_{565}$ sample to that of CdSe$_{494}$; Figure 2b,c summarizes the single-particle results for these two sizes. CdSe$_{565}$ has an ensemble PL line width of 111 meV *versus* 122 meV observed for CdSe$_{494}$. Although the ensemble line widths are similar, single-particle data show a reduction in the average emission FWHM from 98 to 80 meV for the bigger MSNC. Slightly larger particle-to-particle variations are also observed for CdSe$_{565}$, with a standard deviation of the center emission energy of 38 meV compared to 32 meV for CdSe$_{494}$.

The reduced single-particle line width observed for bigger MSNCs is consistent with trends observed for other 3D-confined systems, such as QDs. Indeed, several studies have reported reduced PL line widths for larger NCs.[6,31,37-40] To explain this trend, a computational atomistic approach has previously been used to study the size dependence of exciton–phonon coupling for CdSe NCs.[34] The reorganization energy of highly confined CdSe systems was found to be dominated by exciton coupling to low-frequency acoustic phonon modes localized on the NC surface. As the NC size increases, the wavefunction overlap between excited carriers and surface atoms decreases, leading to a reduction in exciton coupling to these surface modes and, therefore, a reduction in the reorganization energy.

Even though MSNCs likely have a different shape than QDs,[21] they behave comparably in terms of their line widths. Indeed, the MSNCs exhibit a reduction from 98 to 80 meV as the emission energy shifts from 2.43 to 2.17 eV. This reduction is similar to what was extracted for the homogeneous line widths of CdSe QDs using 2D electronic spectroscopy[39] and photon-correlation Fourier spectroscopy.[31] Moreover, our finding that the spectral widths of MSNCs and QDs behave similarly is consistent with *ab initio* studies that predict similar emission line widths for tetrahedral and spherical CdSe NCs.[37]

Besides size, another property that influences line widths is surface passivation.[31,37,39,41] Therefore, we compared the single-particle spectra of CdSe$_{494}$ with two different core/shell samples.[30] The first, CdSe$_{494}$/CdS, was achieved by growing a thin CdS monolayer on top of a CdSe$_{494}$ core. The second, CdSe$_{494}$/Cd$_x$Zn$_{1-x}$S, was obtained by adding a thick Cd$_x$Zn$_{1-x}$S shell. Previous TEM analysis has shown that these CdSe$_{494}$/CdS MSNCs had an average diameter of $3.13 \pm 0.19$ nm,[30] approximately 0.46 nm larger than the core-only CdSe$_{494}$ (equivalent to 0.75 monolayers of CdS).[42] For CdSe$_{494}$/Cd$_x$Zn$_{1-x}$S, the average diameter obtained was $5.16 \pm 0.47$ nm,[30] which is 2.49 nm bigger than the CdSe$_{494}$ core.

**Figure** a,b shows the ensemble PL spectra and decays for CdSe$_{494}$, CdSe$_{494}$/CdS, and CdSe$_{494}$/Cd$_x$Zn$_{1-x}$S. From the PL spectra, we observe a redshift from 2.43 eV for CdSe$_{494}$ to 2.37 and 2.30 eV for CdSe$_{494}$/CdS and CdSe$_{494}$/Cd$_x$Zn$_{1-x}$S, respectively. This shift is caused by the delocalization of the excitonic wavefunction into the shell. From the time-resolved PL of the core-only sample, we see a multi-exponential decay, which is evidence of multiple relaxation pathways for the exciton. For the shelled samples, we observe near-monoexponential traces with decay constants of 23 ns (for CdSe$_{494}$/CdS) and 19 ns (for CdSe$_{494}$/Cd$_x$Zn$_{1-x}$S). We attribute these values to the lifetime of the band-edge recombination. The added shells also lead to a strong increase in PL QY, from 38% for CdSe$_{494}$ to 78 and 80% for CdSe$_{494}$/CdS and CdSe$_{494}$/Cd$_x$Zn$_{1-x}$S, respectively. The increase in the PL QY and the suppression of the multi-exponential decay components are strong indicators of improved surface passivation due to the shelling process.[43]

The ensemble PL spectra exhibit emission line widths of 122 meV for CdSe$_{494}$ and 128 and 117 meV for CdSe$_{494}$/CdS and CdSe$_{494}$/Cd$_x$Zn$_{1-x}$S, respectively (Figure 3a). To investigate the effect of surface passivation on the line widths, we measured the single-particle emission spectra of these samples (Figure 3c,d). Similar to what was observed for the larger CdSe$_{565}$, we see a reduction in the average single-particle emission FWHM, from 98 meV to 89 meV and 91 meV for the CdS and Cd$_x$Zn$_{1-x}$S shells, respectively.

A decrease in line width after surface passivation has previously been reported for QDs.[31,39] However, different interpretations of this effect have been proposed. For example,[31] a reduction of homogeneous line widths has been observed by either growing a shell[31] or changing the passivating ligand[39] in CdSe QDs. In both cases, this reduction was attributed to the suppression of coupling between excitons and optical phonons. The dependence of the QD line width on size and surface passivation was explained by the presence of surface traps that induce an internal electric field in the NC, which enhances exciton–optical-phonon coupling through the Fröhlich interaction.[44-46] In contrast, simulations that investigated the effect of shell growth on CdSe QDs showed that exciton coupling to acoustic and optical phonons was suppressed due to the spatial decoupling of excitons and surface atoms.[34] In that case, reduced coupling to the low-frequency acoustic phonons was found to be primarily responsible for changes in the exciton reorganization energy,[34] which presumably leads to narrower line widths.

To investigate the influence of charging on the line widths, we performed time-resolved single-particle PL experiments for all studied particles. PL decays were taken right after measuring the single-particle PL spectra. In general, the PL intensity of single NCs exhibits intermittency, also known as blinking.[47] This effect has been explained by charging of the NC.[47,48] Therefore, by examining time traces of the PL intensity from individual MSNCs, we can investigate the role of charge on their behavior.

Figure S3 in the Supporting Information shows an example of a PL intensity trace of a single CdSe$_{494}$/Cd$_x$Zn$_{1-x}$S. From this graph, we observe that, as with other NCs,[47] single MSNCs blink. In addition to changing the PL intensity, charging is also expected to alter the decay lifetime due to modifications in the NC environment.[49] This change can be visualized by plotting fluorescence lifetime intensity distribution (FLID) maps.[49] Figure 3e shows an example of a FLID map for one of the individual CdSe$_{494}$/Cd$_x$Zn$_{1-x}$S. We observe that the PL from this MSNC is distributed between two different states in terms of intensity and lifetime. One is brighter with a lifetime centered at 18 ns, while the other is dimmer and centered around 7 ns. We attribute this behavior to a neutral and charged MSNC state, respectively.

To quantify the likelihood of charging in each MSNC, we used such FLID maps to determine the percentage of photons emitted from a charged state in relation to the total emission. Figure 3f shows the acquired line width (FWHM) as a function of the percentage of photons detected from the charged state for all of our studied MSNCs (615 particles). We do not find a substantial correlation between the line widths and the likelihood that the MSNC is charged (Pearson coefficient of 0.12). Therefore, our data do not support the explanation that charge variations between our MSNCs cause differences in their line widths. However, we do see a stronger correlation (Pearson coefficient of 0.42) between the line widths and the emission energy of the particles (Figure S4). This correlation suggests that exciton confinement is responsible (at least partially) for the different line widths observed between samples. Higher confinement enhances overlap between the exciton and the surface atoms, presumably broadening emission due to increased coupling with surface-localized acoustic phonons.[34]

**Comparison With Other CdSe Systems.** In Figure 4, we plot the ensemble line widths of CdSe MSNCs and other CdSe-based NCs as a function of their emission energy. We extracted data from published PL spectra for different sizes, shapes, and shell configurations (see Methods). We observe that for 3D-confined systems (such as core-only and core/shell QDs and MSNCs), the line widths depend strongly on size. For QDs, the samples that exhibit the narrowest ensemble line widths are large (emitting below 2.0 eV). To date, MSNCs with comparable sizes cannot be synthesized. However, the largest MSNCs (*i.e.,* those emitting between 2.2 and 2.4 eV) perform comparably to QDs at the same emission energies. Moreover, the smallest MSNCs (*i.e.,* those emitting above 2.4 eV) actually outperform QDs at the same energy (for the data available). Note that CdSe NPLs have line widths significantly narrower than both QDs and MSNCs. They represent a system with one-dimensional (1D) confinement, for which the extracted FWHM depends much less on the emission energy.

Based on these comparisons, two challenges exist for obtaining NCs with sharp emission line widths and high confinement in three dimensions. The first is related to the single-particle line width. In highly confined systems, the wavefunctions of excited carriers overlap more with surface atoms, leading to enhanced coupling with phonons[34,50] and, therefore, broader single-particle line widths. The second is related to particle-to-particle variations. With stronger confinement, the emission energy is more sensitive to the NC size. Therefore, smaller morphological variations among particles cause larger differences in the emission energy of individual NCs. If particle-to-particle emission variations are eliminated, the ensemble PL line width is determined by the single-particle PL. However, this limit, in which the ensemble and single particles exhibit the same PL line width, has been achieved only for CdSe QDs with weaker confinement (*i.e.*, PL emission energies below 2.1 eV).[12,32,43] Another material that has been reported to have this characteristic is NPLs.[51] In this case, strong 1D confinement is defined by the NPL thickness (which is truly monodisperse), and small changes in the lateral size of the NPLs do not significantly change their emission spectra.[52,53]

**Advantages of Strong 3D Confinement.** While the small sizes of MSNCs are responsible for broadening their line widths, some applications could benefit from the stronger 3D confinement inherent to these nanostructures. For example, to generate entangled photon pairs, high biexciton binding energies (which increase with confinement[54-56]) are desirable.[57] High biexciton binding energies are also beneficial for lowering the lasing threshold due to reduced reabsorption.[58-60] Moreover, confinement enhances single-photon emission purity due to the increased probability of Auger recombination.[9,61]

To exemplify how the small sizes of MSNCs can be used to leverage relevant properties for applications, we investigated the single-photon emission purity of single MSNCs. We measured the second-order correlation function [$g^{(2)}(\tau)$] for $CdSe_{494}/Cd_xZn_{1-x}S$ (which emits near 2.30 eV) and extracted $g^{(2)}(0)$ (see Methods). This parameter signifies the probability for two photons to emit after one excitation pulse. Thus, it directly relates to the single-photon purity of the emitter.[62] Figure 5a shows a histogram with the distribution of our measured $g^{(2)}(0)$ values. The average $g^{(2)}(0)$ is 0.14, and the lowest is 0.05. In Figure 5b, we plot the normalized second-order correlation function for one of our best MSNCs. Our extracted $g^{(2)}(0)$ value is superior to those reported for bigger CdSe QDs[63-68] and comparable to those for smaller CdSe QDs with stronger confinement[69] [$g^{(2)}(0) = 0.04$ for a CdSe/ZnS QD emitting at 2.17 eV]. Our results are also consistent with those obtained for InP QDs[70,71] and $CsPbI_3$ perovskite NCs,[61,72] which are materials with much larger exciton Bohr radii (10 nm for InP[73] and 12 nm for $CsPbI_3$,[74] while in CdSe, it is 5.6 nm[75]). Because of the enhanced confinement, we hypothesize that InP MSNCs should demonstrate an even higher single-photon emission purity than the CdSe MSNCs shown here.

**CONCLUSIONS**

The discrete growth of MSNCs presents a potential strategy for producing truly monodisperse colloidal NCs in 3D. However, it has been puzzling why ensembles of CdSe MSNCs show broader emissions than state-of-the-art CdSe QDs. We used single-particle spectroscopy to explore this issue. Our PL spectra reveal that MSNCs do exhibit particle-to-particle inhomogeneities, leading to modest broadening of the ensemble emission spectra. This effect could be due to structural variations (*e.g.*, missing atoms on edges and vertices) between individual MSNCs. More importantly, our studies indicate that the largest contribution to the ensemble PL line width of MSNCs is due to the single-particle emission (FWHM above 80 meV for the particles studied here). We find that such line widths are consistent with strong exciton–acoustic-phonon coupling. From all of our experiments, we conclude that the line widths of MSNCs are broader than state-of-the-art QDs because MSNCs are significantly smaller. Both broadening mechanisms (particle-to-particle variations and exciton–phonon coupling) are enhanced with increased 3D exciton confinement. Increasing the size of MSNCs or adding a shell suppresses the exciton–phonon coupling and reduces the single-particle line width. However, for stronger confinement regimes (*i.e.,* NCs that emit above 2.4 eV), the line widths of MSNCs are actually sharper than those from the best reported QDs.

Even though the small sizes of MSNCs broaden their emission line widths, this material is a promising alternative to QDs when larger 3D confinement is desired. For example, they allow the emission energy of a material to be shifted to higher energies. Another example where the high confinement present in MSNCs is beneficial is in single-photon emitters. Indeed, the MSNCs studied here demonstrated $g^{(2)}(0)$ values as low as 0.05. Overall, we believe that MSNCs possess a unique set of characteristics. They are colloidally stable materials that are highly emissive, produce strong exciton confinement in 3D, and are potentially monodisperse. We expect them to be particularly promising for applications where all of these characteristics are required.

**METHODS**

**Single-Particle Spectroscopy.** To obtain a film with isolated single particles, the MSNCs were diluted in toluene with 1 wt% Zeonex polymer at a concentration 5 orders of magnitude lower than that of the initial synthesis product. The diluted samples were then spin-coated onto a quartz slide for 1 min at 2000 revolutions per min. For optical measurements, the samples were excited with a 405 nm pulsed diode laser (Picoquant, LDH-D-C-405) with a 1 MHz repetition rate. The excitation intensity was kept between 1.5 and 3.1 μW cm$^{-2}$ in all experiments. The emission was collected with an oil-immersion objective (Nikon, SR HP Apo TIRF), with 100× magnification and 1.49 numerical aperture. Single-particle experiments were automated to measure multiple particles. For this, a micro-PL map was first collected to determine single-NC positions by scanning a 60 μm$^2$ region with piezo stages (Mad City Labs, NANO-LPS300) while directing the filtered PL to an avalanche photodiode (APD; Excelitas, SPCM-AQRH-14-TR). After acquiring the positions, the piezo stages moved to each NC location in series to acquire its PL spectrum, followed by its time-resolved PL. To switch between the two measurements, an automatic flip mirror was deployed. For the spectral measurements, the emission was directed to an imaging spectrometer (Andor, Shamrock 303i), dispersed with a grating (300 lines mm$^{-1}$), and imaged with an electron-multiplying charged-coupled device (EMCCD) camera (Andor, iXon 888 Ultra). For the time-resolved PL measurements, the emission was sent to a Hanbury-Brown–Twiss (HBT) setup with a 50/50 non-polarizing beam splitter and two APDs (Excelitas, SPCM-AQRH-14-TR). The APDs were connected to a time-tagger box (Picoquant, HydraHarp 400) for time-correlated single-photon counting (TCSPC). The HBT setup was used to measure the FLID maps and the second-order correlation functions.

**Extraction of $g^{(2)}(0)$.** To extract $g^{(2)}(0)$, we fitted the normalized second-order correlation function with mono-exponential decays, following a method previously reported.[61] We defined $g^{(2)}(0)$ as the ratio of the integrated area of the fitted peak located at $\tau = 0$ divided by the average area of the 6 closest peaks. No spectral filtering was used to block biexciton emission, and no background correction was used to extract $g^{(2)}(0)$.

**Synthesis of the Studied MSNCs.** Unless noted otherwise, all commercially obtained reagents/solvents were used as received. Cadmium oxide (#48-0800, 99.999% Cd) and cadmium acetate (#48-0100, 99.999%) were purchased from Strem Chemicals. Trifluoroacetic acid (#302031, ≥99%), trifluoroacetic anhydride (#106232, ≥99%), oleic acid (#364525, 90%), dichloromethane (DCM, #676853, ≥99.5%), acetonitrile (MeCN, #34998, ≥99.9%), zinc oxide (#255750, 99.99%), hexadecane (#H6703, 99%), methyl acetate (MeOAc, #W267600, ≥98%), hexane (#34859, ≥97%), and selenium (#204307, ≥99.999%) were purchased from Sigma-Aldrich. Triethylamine (#157910010, 99%) was purchased from Acros Organics. Deionized (DI) water was obtained from a MilliQ Advantage A10 water purification system (Merck Millipore). Bis(stearoyl) selenide was synthesized according to a literature procedure,[21] as were cadmium oleate, CdSe$_{494}$/CdS, and CdSe$_{494}$/Cd$_x$Zn$_{1-x}$S.[30]

For the synthesis of CdSe$_{494}$, cadmium oleate (540.2 mg, 0.8 mmol) and cadmium chloride (73.3 mg, 0.4 mmol) were added to a 50 mL three-neck round-bottom flask and suspended in 20 mL hexadecane. The mixture was heated to 110 °C and degassed under vacuum for 30 min. The mixture was brought back under N$_2$ and heated to 240 °C. At the same time, bis(stearoyl) selenide (122.8 mg, 0.2 mmol) was dissolved in 2.5 mL dry toluene under N$_2$. Mild heating (~35 °C) was used to completely dissolve the selenide. Once the cadmium oleate solution reached 180 °C, the bis(stearoyl) selenide solution was rapidly injected. The MSNCs were allowed to grow at 180 °C. After 50 min, the flask was rapidly cooled to room temperature with a water bath. At room temperature, the ~22 mL of crude material was split into two 50 mL centrifuge tubes. 14 mL MeOAc was added to each tube. After waiting 5 min, the tubes were centrifuged at 8586 g (8000 rpm) for 5 min. The off-white precipitate was discarded while the supernatant was transferred to separate centrifuge tubes. Another 5 mL MeOAC was added to this supernatant. After 5 min, these tubes were centrifuged at 8586 g (8000 rpm) for 5 min. The white precipitate was again discarded, and the supernatant was transferred to separate centrifuge tubes. Another 10 mL MeOAC was added to this supernatant, and these tubes were centrifuged at 8586 g (8000 rpm) for 5 min. The supernatant was discarded, and the precipitate in each tube was redispersed with 6 mL hexane. In some syntheses, an insoluble gel remained in the centrifuge tubes and was discarded. These two portions were combined in a new centrifuge tube. 18 mL MeOAc was added to this dispersion, then centrifuged at 8586 g (8000 rpm) for 5 min. The supernatant was transferred to a new centrifuge tube while the white precipitate was discarded. A final 12 mL MeOAc was added to the supernatant and centrifuged at 8586 g (8000 rpm) for 5 min. The supernatant was discarded, and the precipitate was redispersed in 2 mL hexane, yielding $CdSe_{494}$. The solution was stored in a vial in ambient conditions and was colloidally stable for months.

For the synthesis of $CdSe_{565}$, cadmium oleate (540.2 mg, 0.8 mmol) and cadmium chloride (73.3 mg, 0.4 mmol) were added to a 50 mL three-neck round-bottom flask and suspended in 20 mL hexadecane. The mixture was heated to 110 °C and degassed under vacuum for 30 min. The mixture was brought back under $N_2$ and heated to 240 °C. At the same time, bis(stearoyl) selenide (122.8 mg, 0.2 mmol) was dissolved in 2.5 mL dry toluene under $N_2$. Mild heating (~35 °C) was used to dissolve the selenide completely. Once the cadmium oleate solution reached 240 °C, the bis(stearoyl) selenide solution was rapidly injected. The system was manually purged of toluene to return the temperature to 240 °C in ~10 min. This was done by using a 1 mL syringe to pull out 1 mL of gas from the reaction flask every 2 min for 10 min. The MSNCs were allowed to grow at 240 °C. After 90 min, the flask was rapidly cooled to 180 °C with an air gun, then to room temperature with a water bath. At room temperature, 5 mL hexane was added, and the reaction mixture was centrifuged at 8586 g (8000 rpm) for 5 min. The supernatant was carefully decanted (~15 mL total volume), and 25 mL MeOAc was added to this supernatant. This mixture was centrifuged at 8586 g (8000 rpm) for 5 min. The supernatant was carefully decanted, and 10 mL MeOAc was added to this supernatant. This mixture was centrifuged at 8586 g (8000 rpm) for 5 min. The supernatant was discarded, and the precipitate was redispersed in 5 mL hexane. 8 mL MeOAc was added to this dispersion, and the mixture was centrifuged at 8586 g (8000 rpm) for 5 min. The supernatant was discarded, and the precipitate was redispersed in 1 mL hexane. This dispersion was centrifuged at 8586 g (8000 rpm) for 5 min, and the supernatant was isolated, yielding $CdSe_{565}$. The solution was stored in a vial in ambient conditions and was colloidally stable for months.

**Ensemble Optical Spectroscopy.** Optical properties for ensemble samples were taken with the MSNCs dispersed in hexane in quartz cuvettes. PL spectra and decays were acquired using an Edinburgh Instruments FLS 980 fluorometer. The absorption spectra were measured with a Varian Cary Scan 50 spectrophotometer.

**Comparison of MSNCs with Other CdSe-Based NC Systems.** For a fair comparison, all of the literature results shown in Figure 4 were analyzed using the same procedure. The data were extracted from the published PL spectra[12,19,21,30-33,43,76-84] using the WebPlotDigitilizer tool[85] and fitted with a Gaussian function.

ASSOCIATED CONTENT

**Supporting Information.**

Validation of single-particle spectroscopy experiments and further data supporting the conclusions in the main text.

AUTHOR INFORMATION

**Corresponding Author**

*Email: dnorris@ethz.ch

**ORCID**

Gabriel Nagamine: 0000-0002-4830-7357

Julian Santen: 0009-0005-3353-1769

Juri G. Crimmann: 0000-0002-0367-5172

Aniket S. Mule: 0000-0001-8387-080X

Andrew B. Pun: 0000-0002-3052-912X

David J. Norris: 0000-0002-3765-0678

**Note**

The authors declare no competing financial interest.

ACKNOWLEDGMENTS

This work was supported by the Swiss National Science Foundation (SNSF) under Award No. 200021_188593. The authors thank D. Petter, J. J. E. Maris, R. Keitel, and A. Cocina for helpful discussions and M. Haug for technical assistance.

**Table 1.** Comparison of PL line widths from CdSe MSNCs with other CdSe-based nanocrystals.

| System | FWHM (meV) | Reference |
|---|---|---|
| MSNCs | 129–169 | 21 |
| Core-only QDs | 110–140 | 31 |
| Core/CdS shell QDs | 68–96 | 32 |
| Asymmetrically-strained core/$Cd_xZn_{1-x}Se$ QDs | 60–76 | 33 |
| Nanoplatelets | 37–78 | 19 |

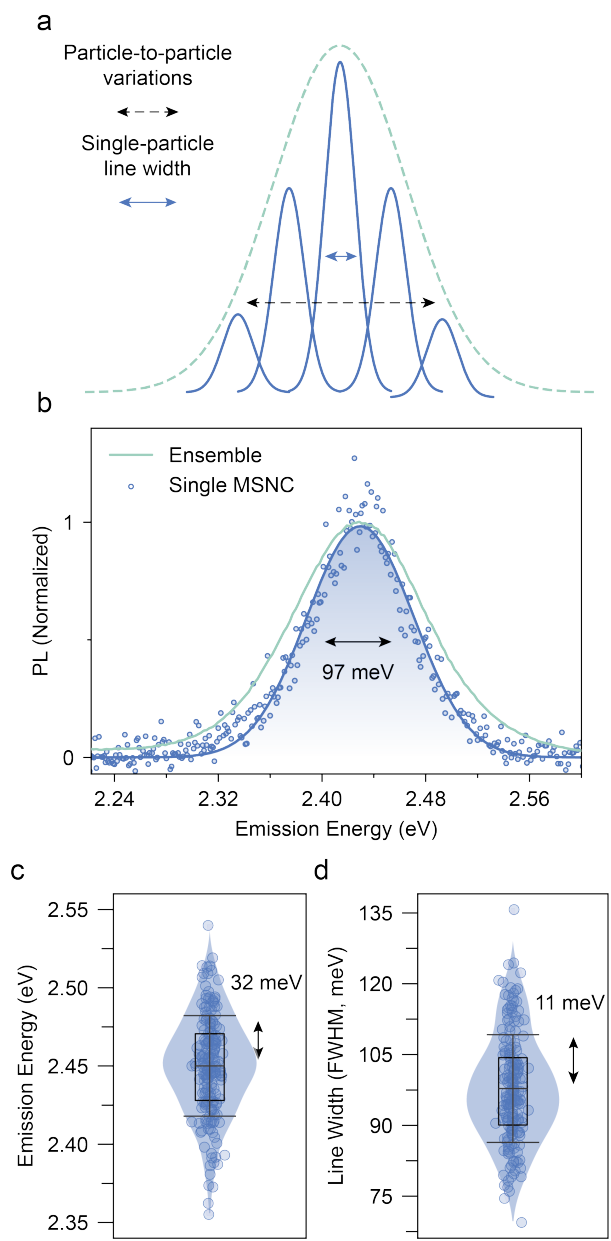

**Figure 1.** Emission broadening mechanisms in magic-sized nanocrystals (MSNCs). (a) Schematic of the ensemble photoluminescence (PL) spectrum (green dashed line), which contains contributions from the single-particle line width (blue Gaussian curves) and particle-to-particle variations (black dashed arrow). (b) A comparison of the measured ensemble PL spectrum (green solid line) with that from a representative single MSNC (blue scattered circles) for $CdSe_{494}$ at room temperature. The solid blue line shows a Gaussian fit of the single-MSNC PL spectrum. (c, d) Statistics for the measured PL energies and line widths, respectively, for many individual $CdSe_{494}$. The parameters were obtained by fitting the emission spectra with Gaussian curves. The box delimits the 25th to 75th percentiles of the data, and the horizontal lines denote one standard deviation from the mean (center line). The distribution density of the data is represented with blue shading (violin plot).

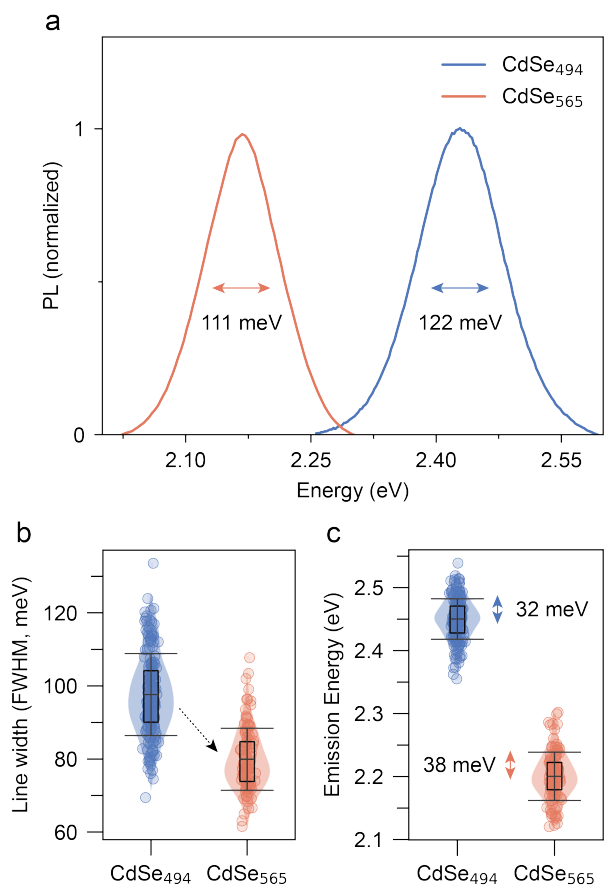

**Figure 2.** Size-dependent line widths of CdSe MSNCs. (a) Ensemble PL emission spectra from CdSe$_{565}$ (orange curve) and CdSe$_{494}$ (blue curve), measured at room temperature. (b, c) Statistics, plotted as in Figure 1c,d for the measured PL line widths and energies, respectively, for many individual CdSe$_{494}$ and CdSe$_{565}$.

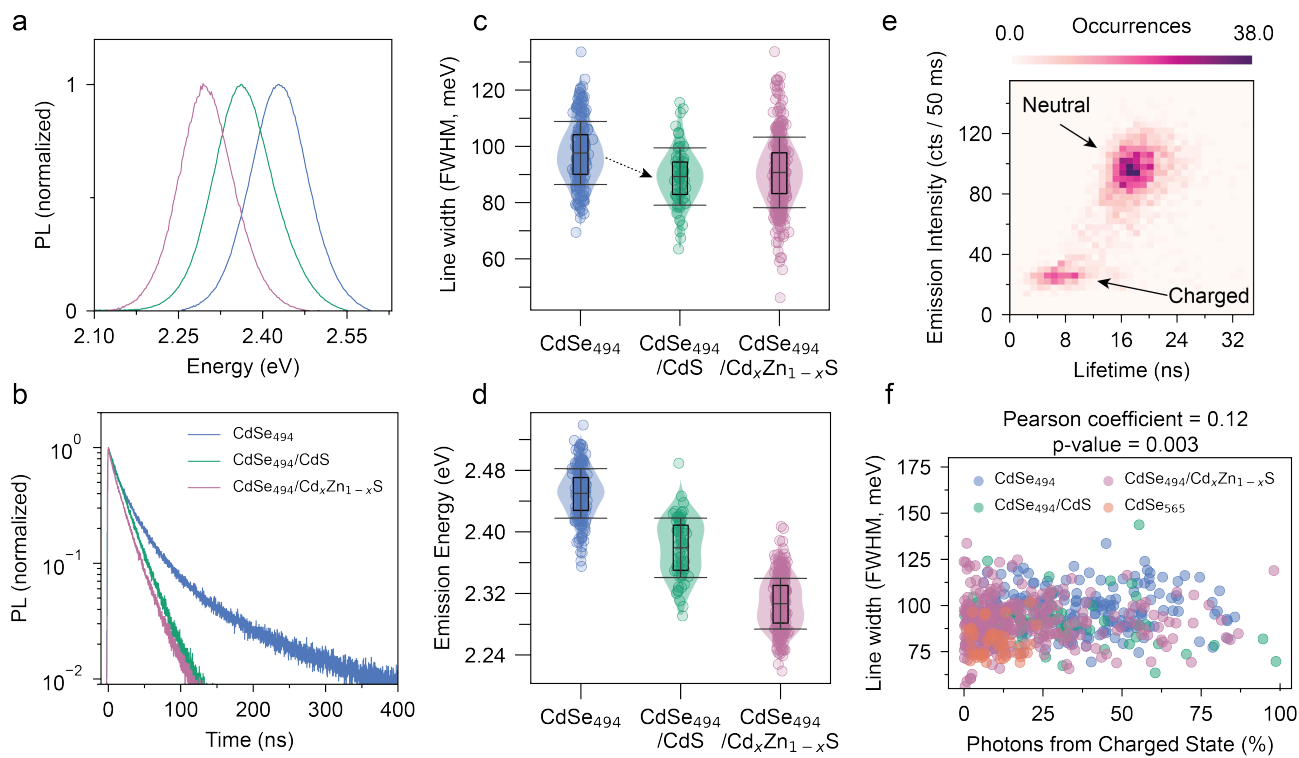

**Figure 3.** The influence of surface passivation on the PL line widths of CdSe MSNCs. (a) Spectral and (b) time-resolved ensemble PL measurements on CdSe$_{494}$ (blue curves), CdSe$_{494}$/CdS (green curve), and CdSe$_{494}$/Cd$_x$Zn$_{1-x}$S (purple curves). Measurements were taken at room temperature for samples diluted in hexane. (c, d) Statistics, plotted as in Figure 1c,d, for the measured PL line widths and energies, respectively, for many individual CdSe$_{494}$, CdSe$_{494}$/CdS, and CdSe$_{494}$/Cd$_x$Zn$_{1-x}$S. (e) Example of a fluorescence lifetime intensity distribution (FLID) map for an individual CdSe$_{494}$/Cd$_x$Zn$_{1-x}$S. (f) Measured line widths as a function of the percentage of photons emitted from a charged state (Pearson coefficient of 0.12, p-value of 0.003).

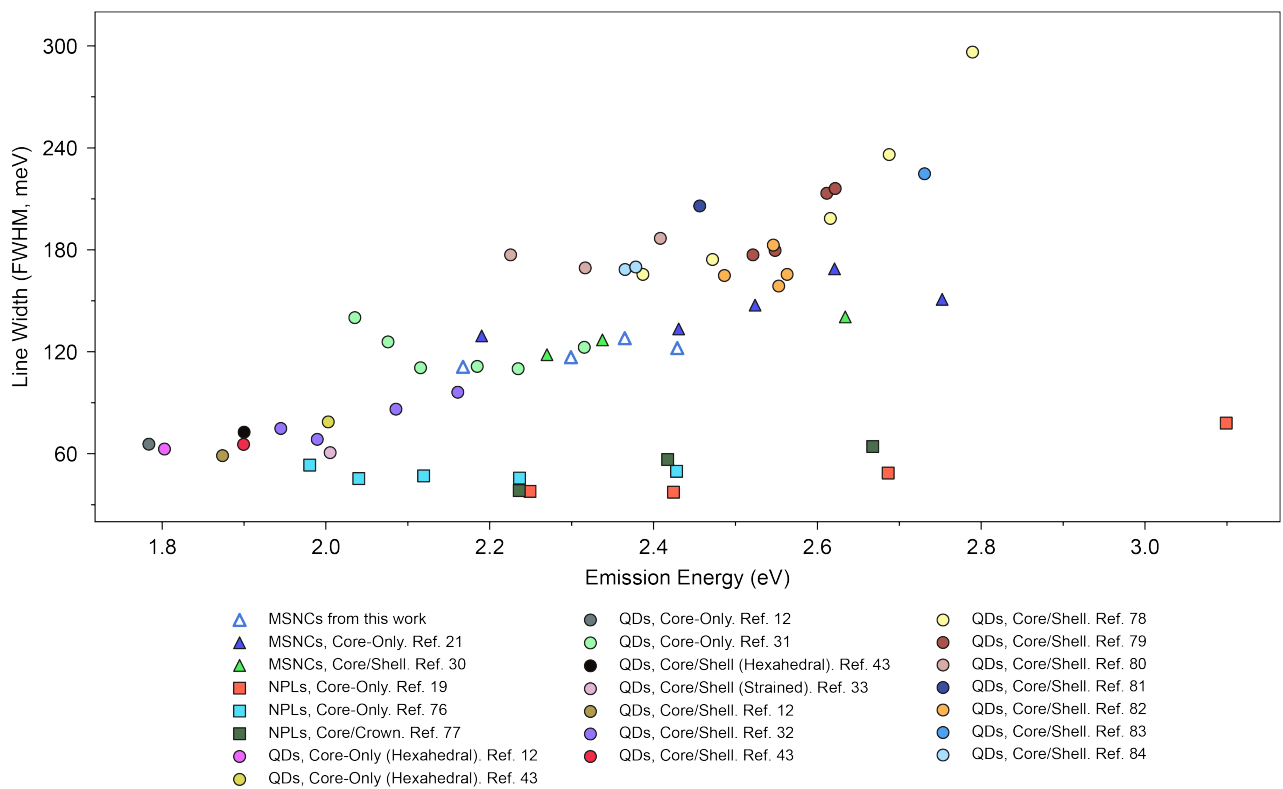

**Figure 4.** Comparison of the optical properties of CdSe MSNCs with other CdSe-based nanocrystals reported in the literature. The ensemble room-temperature PL line width (FWHM) is plotted *versus* emission energy. Data from the literature[12,19,21,30-33,43,76-84] was extracted using the WebPlotDigitizer tool,[85] and the same fitting procedure as in Figure 1c,d was applied. The values shown in this plot are also provided in Table S2 in the Supporting Information.

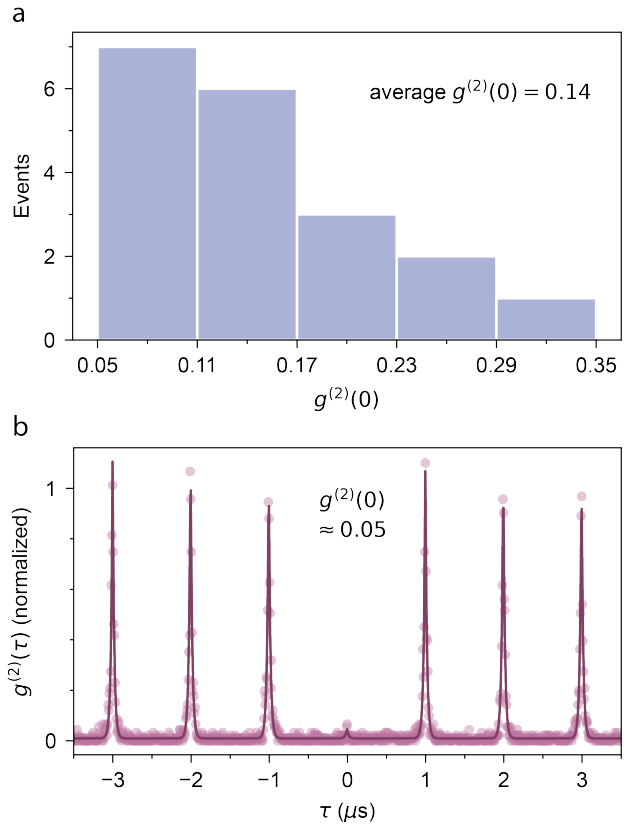

**Figure 5.** Single-photon emission purity of $CdSe_{494}/Cd_xZn_{1-x}S$ at room temperature. (a) Histogram of the measured normalized second-order correlation function at a delay time of $\tau = 0$, denoted as $g^{(2)}(0)$. (b) Measured normalized $g^{(2)}(\tau)$ for the individual $CdSe_{494}/Cd_xZn_{1-x}S$ with the lowest $g^{(2)}(0) = 0.05$. The fitting (solid line) was performed following a method previously reported.[61]

**Table of Contents Graphic**

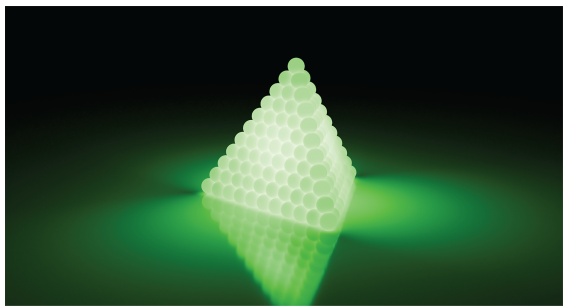

Supporting Information for

# Spectroscopy of Single CdSe Magic-Sized Nanocrystals

*Gabriel Nagamine, Julian Santen, Juri G. Crimmann, Aniket S. Mule, Andrew B. Pun,*

*and David J. Norris*[*]

Optical Materials Engineering Laboratory, Department of Mechanical and Process Engineering,

ETH Zurich, 8092 Zurich, Switzerland

**S1. Validation of Single-Particle Spectroscopy Experiments**

In the single-particle spectroscopy experiments, an automated setup was used to avoid user bias and increase the number of measured single MSNCs (see Methods). In Figure S2, we plot the normalized sum of all single-particle spectra together with their ensemble spectra. This plot is helpful for checking the representability of the single-particle data (compared to the ensemble) and determining if photodegradation processes are affecting the line widths of the particles during the single-particle experiments. The spectra obtained by summing the single-particle spectra are consistent with the ensemble measurements. We observe a small blue shift of 14 and 27 meV for $CdSe_{494}$ and $CdSe_{565}$, respectively, which is consistent with photooxidation[S1,S2] in some of the studied particles. The line widths extracted in the spectra obtained by summing the single-particles are very similar to the ensemble. A summary with the fitting parameters obtained by analyzing the ensemble and summed single-particle spectra can be found in Table S1.

## S2. Supporting Figures

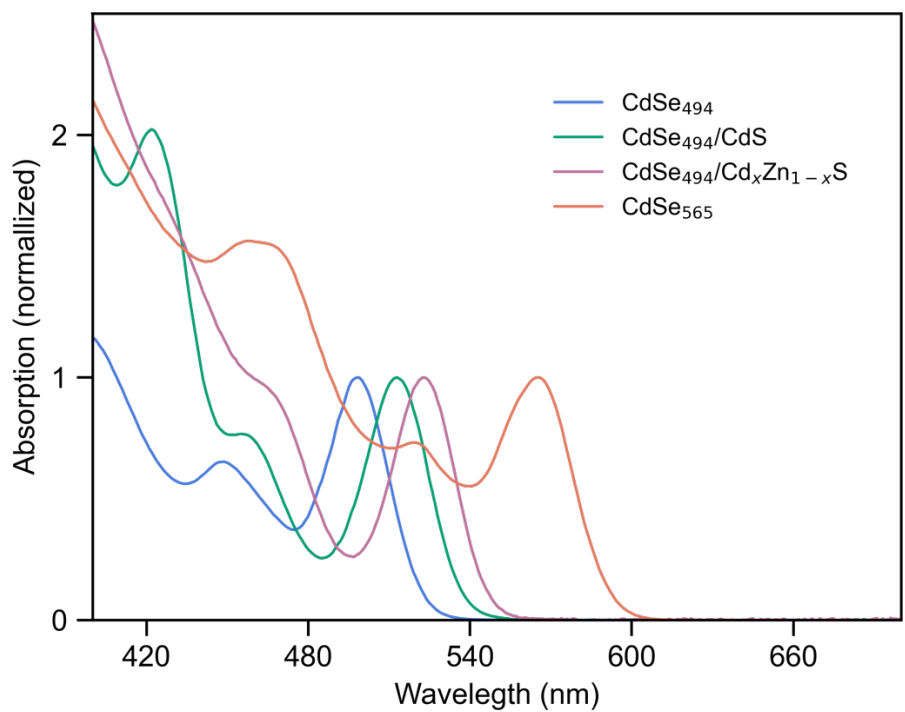

**Figure S1.** Absorption spectra of the CdSe$_{494}$, CdSe$_{494}$/CdS, CdSe/Cd$_x$Zn$_{1-x}$S, and CdSe$_{565}$ samples. Measurements were taken at room temperature with the samples diluted in hexane.

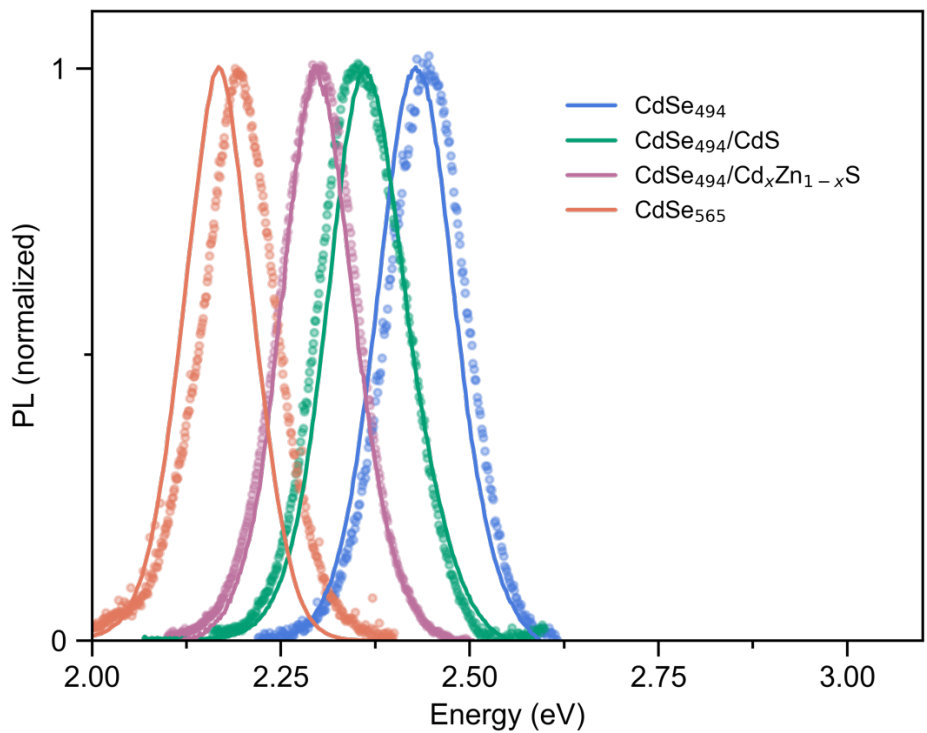

**Figure S2.** Validation of the representability of the single-particle experiments. Comparison of the ensemble photoluminescence spectra (solid lines) with the spectra obtained by summing all measured single-particle spectra (circles). The ensemble spectra were obtained at room temperature with the MSNCs diluted in hexane.

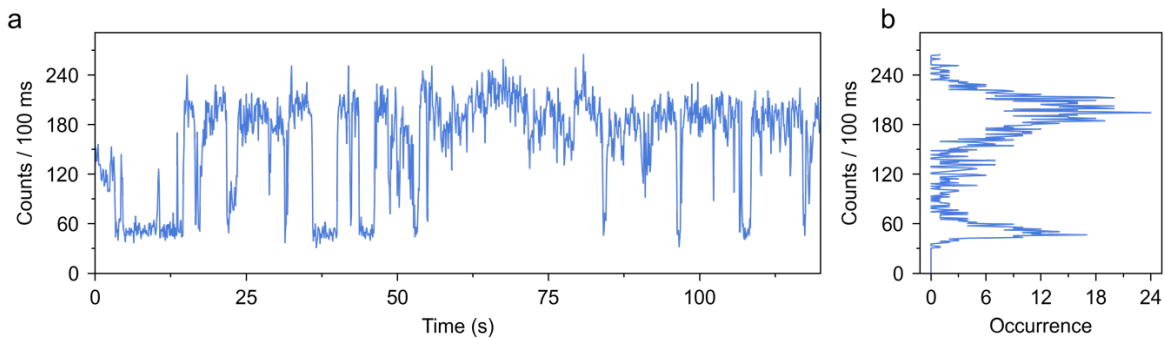

**Figure S3.** An example intensity trace for one of the studied CdSe/Cd$_x$Zn$_{1-x}$S magic-sized nanocrystals. (a) Time-resolved trajectory of 2 min of single-particle emission. (b) Histogram of the intensity distribution during this period.

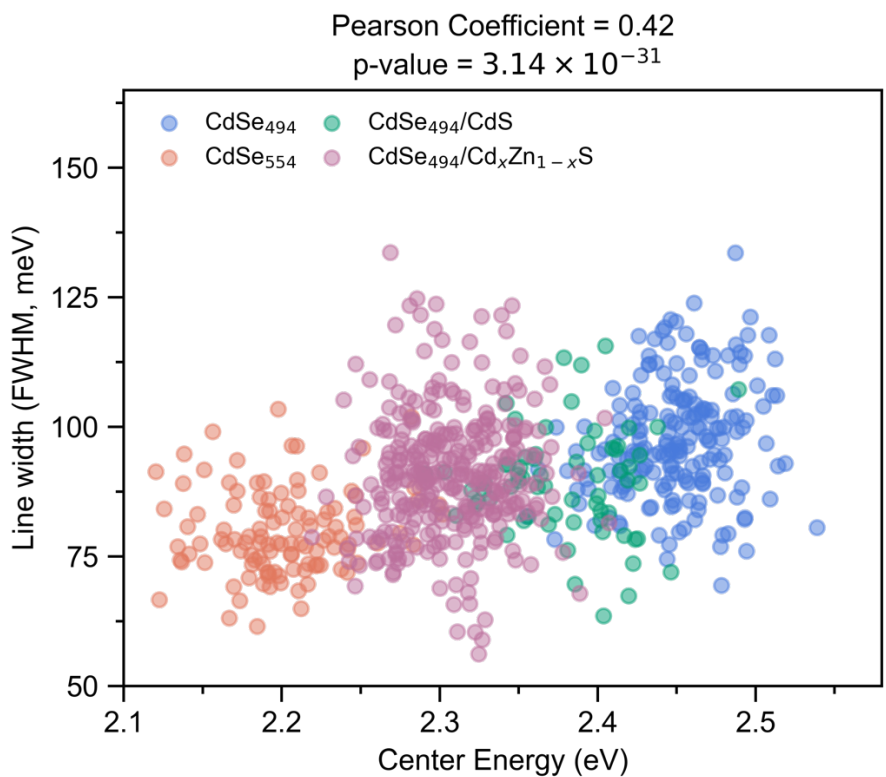

**Figure S4.** Correlation between emission line widths and energies for all studied MSNCs. The Pearson coefficient between emission line width and energy was 0.42, with a p-value of $3.14 \times 10^{-31}$.

**S3. Supporting Tables**

**Table S1.** Comparison of PL line widths and center emission energies from ensemble measurements and by summing single-MSNC spectra.

| Sample | Ensemble Line Width (FWHM, meV) | Summed Single-Particle Line Width (FWHM, meV) | Ensemble Center Energy (eV) | Summed Single-Particle Center Energy (eV) |
|---|---|---|---|---|
| $CdSe_{494}$ | 122 | 123 | 2.43 | 2.44 |
| $CdSe_{494}/CdS$ | 128 | 134 | 2.37 | 2.36 |
| $CdSe_{494}/Cd_xZn_{1-x}Se$ | 117 | 123 | 2.30 | 2.30 |
| $CdSe_{565}$ | 111 | 123 | 2.17 | 2.19 |

**Table S2.** Line width and emission energies extracted from published PL spectra of CdSe-based nanocrystals (data plotted in Figure 4 of the main text). References cited are from the main text.

| NC Structure and Reference | Line Width Range (FWHM, meV) | Emission Energy Range (meV) |
|---|---|---|
| MSNCs, Core-Only. Ref. 21 | 129.0 – 169.0 | 2.19 – 2.75 |
| MSNCs, Core/Shell. Ref. 30 | 118.0 – 141.0 | 2.27 – 2.63 |
| NPLs, Core-Only. Ref. 19 | 37.0 – 78.0 | 2.25 – 3.1 |
| NPLs, Core-Only. Ref. 76 | 45.0 – 53.0 | 1.98 – 2.43 |
| NPLs, Core/Crown. Ref. 77 | 38.0 – 64.0 | 2.24 – 2.67 |
| QDs, Core-Only (Hexahedral). Ref. 12 | 63.0 | 1.8 |
| QDs, Core-Only (Hexahedral). Ref. 43 | 79.0 | 2.0 |
| QDs, Core-Only. Ref. 12 | 66.0 | 1.78 |
| QDs, Core-Only. Ref. 31 | 110.0 – 140.0 | 2.04 – 2.32 |
| QDs, Core/Shell (Hexahedral). Ref. 43 | 73.0 | 1.9 |
| QDs, Core/Shell (Strained). Ref. 33 | 61.0 | 2.01 |
| QDs, Core/Shell. Ref. 12 | 59.0 | 1.87 |
| QDs, Core/Shell. Ref. 32 | 68.0 – 96.0 | 1.94 – 2.16 |
| QDs, Core/Shell. Ref. 43 | 66.0 | 1.9 |
| QDs, Core/Shell. Ref. 78 | 166.0 – 296.0 | 2.39 – 2.79 |
| QDs, Core/Shell. Ref. 79 | 177.0 – 216.0 | 2.52 – 2.62 |
| QDs, Core/Shell. Ref. 80 | 169.0 – 581.0 | 2.23 – 2.44 |
| QDs, Core/Shell. Ref. 81 | 206.0 | 2.46 |
| QDs, Core/Shell. Ref. 82 | 159.0 – 183.0 | 2.49 – 2.56 |
| QDs, Core/Shell. Ref. 83 | 225.0 | 2.73 |
| QDs, Core/Shell. Ref. 84 | 168.0 – 170.0 | 2.37 – 2.38 |

SUPPLEMENTARY REFERENCES